# Title
# Heart Rate Variability and Respiration Signal as Diagnostic Tools for Late Onset Sepsis in Neonatal Intensive Care Units

Authors
**Yuan Wang**[1,2,5§], **Guy Carrault**[1,2], **Alain Beuchee**[1,2,3], **Nathalie Costet**[1,2], **Huazhong Shu**[4,5], **Lotfi Senhadji**[1,2,5]

[1] INSERM, UMR 1099, Rennes, 35042, France.
[2] Université de Rennes 1, LTSI, Rennes, 35042, France.
[3] CHU Rennes, Pôle Médico-Chirurgical de Pédiatrie et de Génétique Clinique, Rennes, F-35000, France.
[4] School of Computer Science and Engineering, Southeast University, Nanjing, 210096, P. R. China.
[5] Centre de Recherche en Information Biomédicale sino-français (CRIBs).

Email : yuanwang.research@gmail.com

Abstract
Apnea-bradycardia is one of the major clinical early indicators of late-onset sepsis occurring in approximately 7% to 10% of all neonates and in more than 25% of very low birth weight infants in NICU. The objective of this paper was to determine if HRV, respiration and their relationships help to diagnose infection in premature infants via non-invasive ways in NICU. Therefore, we implement Mono-Channel (MC) and Bi-Channel (BC) Analysis in two groups: sepsis (S) vs. non-sepsis (NS). Firstly, we studied RR series not only by linear methods: time domain and frequency domain, but also by non-linear methods: chaos theory and information theory. The results show that alpha Slow, alpha Fast and Sample Entropy are significant parameters to distinguish S from NS. Secondly, the question about the functional coupling of HRV and nasal respiration is addressed. Local linear correlation coefficient $r^2_{t,f}$ has been explored, while non-linear regression coefficient $h^2$ was calculated in two directions. It is obvious that $r^2_{t,f}$ within the third frequency band ($0.2<f<0.4\ Hz$) and $h^2$ in two directions were complementary approaches to diagnose sepsis. Thirdly, feasibility study is carried out on the candidate parameters selected from MC and BC respectively. We discovered that the proposed test based on optimal fusion of 6 features shows good performance with the largest AUC and a reduced probability of false alarm ($P_{FA}$).

---

[§] Dr. Yuan Wang is currently working at University of California, San Francisco (UCSF).

# 1. Introduction

Late-onset sepsis, defined as a systemic infection in neonates older than 3 days, occurs in approximately 7% to 10% of all neonates and in more than 25% of very low birth weight newborns who are hospitalized in Neonatal Intensive Care Units (NICU). The clinical manifestations of neonatal sepsis, whatever the source of infection, are ususally not so evident. Accordingly, lacking in early and adapted interventions always leads to life risk. Therefore, this disease is a major problem resulting in high morbidity and mortality for premature newborns (Philip, 1990).

As we know, sick preterm infants do not show any fever and consequently the possible signs of sepsis may be detected only with blood culture. However, on one hand, the hematological and biochemical markers which have been used in this symptom, not only require invasive procedures which should not be frequently repeated, but also have low predictive values in the early phase of sepsis. On the other hand, it has been observed experimentally that phenomena of apnea-bradycardia happened more constantly in sepsis preterm infants than in non-sepsis ones (Cao *et al.*, 2004).

Apnea is defined as the cessation of breathing for more than 20 seconds, while Bradycardia in preterm infants is defined as a fall in heart rate under 100 beats per minute. Apnea-Bradycardia (AB) episodes are dangerous for preterm infants mainly in 3 aspects:

- It requires invasive resuscitation techniques for the babies,
- It also needs prolongation of hospital stay and then implies extra costs,
- It has neurological impairment during childhood.

Neonate Intensive Care Unit, where ECG and respiration are continuously recorded, provides environment to assist and monitor the development of the newborns. They are regularly equipped with computerized system triggering alarms if vital signs are abnormal. Several interesting clinical results were recently reported by our group including: i) early prediction of bradycardia (Pravisani *et al.*, 2003), (Cruz *et al.*, 2006), ii) early detection of sepsis (Beuchée *et al.*, 2009), (Billois *et al.*, 2012 ) or iii) study of post-immunization effects (Mialet-Marty *et al.*, 2013). Especially, the second topic showed that these neonatal changes in behavior of physiological signals could be used to diagnose sepsis in sick premature infants.

The objective of this paper is to go ahead and to propose a scheme based upon heart rate variability (HRV), respiration and their relationships to diagnose infection in premature infants via non-invasive ways in NICU. In order to reach this goal, three kinds of analysis were conducted between two selected groups of premature infants: sepsis (S) vs. non-sepsis (NS):
1. Mono-Channel (MC) Analysis where only the RR series is considered.
2. Bi-Channel (BC) Analysis where both RR series and respiration are studied.
3. The combination of these two approaches through a unified data fusion framework.

The remainder of this paper is organized as follows: after introduction, section 2.1 presents the medical concept of premature newborns and clinical problems associated with prematurity, as well as the Mono-Channel Analysis. Next, the Inter-relationships between RR series and Respiration are described in section 2.2. The following section 2.3 proposes a new framework of Feasibility Study based on the significant features from Mono-Channel and Bi-Channel analyses. Around these 3 sub-topics, literature reviews are addressed respectively. Thirdly, results are demonstrated in section 3, with the same architecture in order. Fourthly, we discuss all results part by part in section 4. Finally, section 5 briefly summarizes our research work and draws conclusions.

# 2. Methods

## 2.1. The Mono-Channel Analysis using RR series

The heart rate variability (HRV) analysis in neonatology is a useful tool to understand the cardiovascular control system behavior of late-onset sepsis in premature newborns. Starting from the obvious increase in apnea-bradycardia crisis linked with the state of sickness, a way to evaluate the relationship between the infection and its manifestation was already explored. In particular, since apnea-bradycardia was an indication of altered mechanisms of cardiovascular regulation, the HRV investigation on these subjects is an immediately consequent decision. The paper of Beuchée *et al.* (2009) presents the classic methods related to physiological factors, such as RR series distribution patterns ---- Mean, Median, Variance, Skewness, Kurtosis, Sample Asymmetry (SpAs), magnitude of variability in time domain (RMSSD, SD), linear estimates in frequency domain (power of VLF, LF, HF), fractal exponents ($\alpha_{fast}$ and $\alpha_{slow}$) and complexity measurements (Approximate Entropy and Sample Entropy). Methods for estimation of the entropy of a system represented by a time series are well suited to analyse data sets encountered in cardiovascular studies. Approximate entropy (AppEn) is easily applied to clinical cardiovascular time series, yet lead to inconsistent results. Richman and Moorman (2000) developed a new and relevant complexity measure, Sample entropy (SamEn), and have compared AppEn and SamEn by using them to analyze sets of random numbers with known probabilistic character. SamEn agreed with theory much more closely than AppEn over a broad range of conditions. Abnormal heart rate characteristics of reduced variability and transient decelerations are present early in the course of neonatal sepsis. To examine the dynamics, Lake *et al.* (2002) calculated SamEn, a similar but less biased measure than AppEn. The major findings are that entropy falls before clinical signs of neonatal sepsis and that missing points are well tolerated. Richman *et al.* (2004) also proposed closed form estimates of the variance of SamEn.

Cao *et al.* (2004) invented statistical methods for determining stationary of HR data based on the two-sample Kolmogorov–Smirnov (KS) test, and concluded that neonatal HR data cannot be assumed to be stationary, and become even less stationary prior to sepsis.

The literature review above showed that the sepsis has a direct influence on a number of bradycardias and then disrupt the HRV. The purpose of Mono-Channel Analysis proposed here is to look for the best features, which are able to distinguish sepsis from non-sepsis. To fulfill this goal, we probe into both linear and non-linear methods for HRV analysis, and then compare all of these methods in order to find the candidate ways to discriminate between infected and non-infected premature newborns. In this study, we complete with other complexity measurements such as Permutation Entropy (PermEn) (Takens, 1980) (Bandt *et al.*, 2002) and Regularity (Regul) (Porta *et al.*, 1998) (Porta *et al.*, 2000). These entropies are usually carried by cardiovascular signals. The occurrence of disease is associated with a decrease of Entropy, but an increase of Regul.

Figure 1 illustrates the process of experiments in Mono-Channel Analysis. Firstly, we detect RR series from ECG signals. Secondly, we compute features of RR series in windows. Finally, we do statistical analysis to classify sepsis and non-sepsis. The whole process is implemented based on HRV in real time.

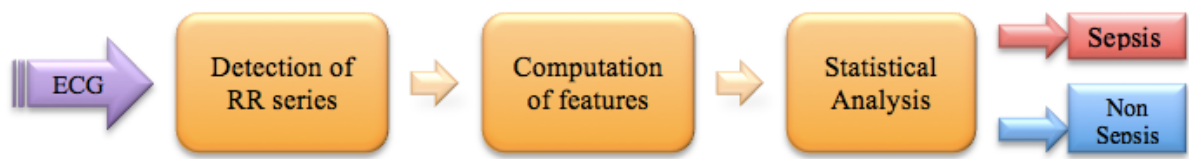

**Figure 1.** Process of experiments in Mono-Channel Analysis.

*2.2. Inter-relationships between RR series and Respiration*

The second part of our research work is concerned with the inter-relationships between RR series and Respiration, which can be regarded as Bi-Channel Analysis. Different mechanisms are involved in the generation of cardiovascular variability rhythms, which have been extensively studied as markers of the sympathovagal interaction controlling cardiovascular functions. Therefore, the methods of multichannel signal analysis can extract more information than it can obtain by the routine techniques of single-channel analysis for HRV signals (Baselli *et al.*, 1994). This is why the Mono-channel signal approaches used to analyze heart rate variability have been extended to several Bi-channel approaches with respect to cardiorespiratory coordination.

Physiologists had already investigated cardiorespiratory coordination in the human organism as early as the 1960's. Calculating the distance between an inspiratory onset and its preceding R-peak, they found intermittent coordination between heartbeat and respiration. In the 1970's this interesting topic was no longer followed up, presumably because the physiological interpretation of the results was limited, although the last reviews of this period appeared in the late 1980's (Raschke, 1987) (Raschke, 1991). The investigation of cardiorespiratory coordination has recently been revived mainly by physicists and mathematicians (Cysarz *et al.*, 2004).

The connections between biological control systems can be revealed sometimes by the presence (in terms of concentration in blood) of mediator elements that can be found during infection manifestation. They are vectors used by the control system, helping to regulate, modulate, and express the body's response to some internal or external perturbation in the system.

In the examined case, sepsis constitutes the perturbation to the system and the role of the Interleukin 1 (Il-1) as vector was dug out. It is one of the first fever effectors in newborns. Some studies have given certain results regarding the role of Il-1 in the connection between infection and respiratory control system: coinciding increase in Interleukin concentration and apnea's crisis was found.

On the other side, instead, no evidence is still present for the relationship between infection and cardiac pacemaker through the Autonomic Nervous System (ANS, see figure 2). The only evidence up to now is that there is an increase in bradycardias together with apneas, during sepsis manifestation.

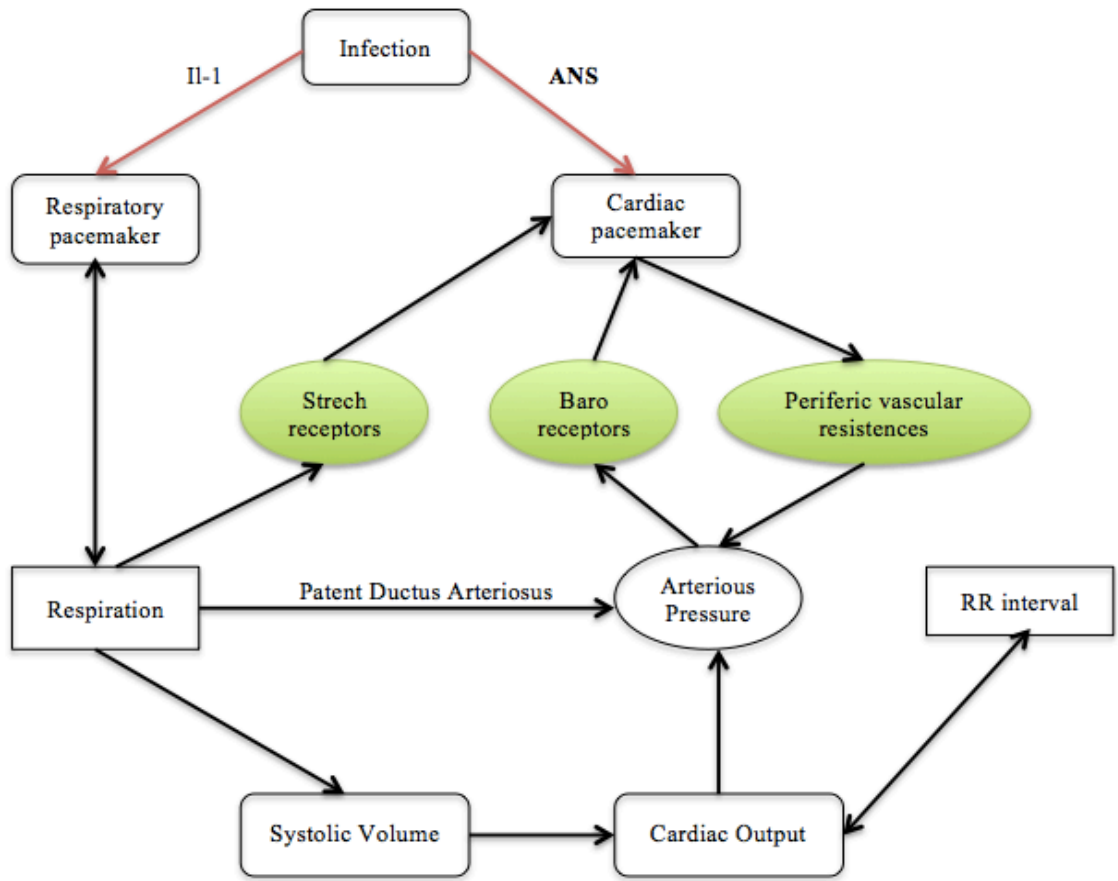

**Figure 2.** Interconnections between cardiovascular systems and respiration. Bi-channel signal analysis rule in their interpretation, to find the infection – bradycardias possible relation.

Of course, giving an answer to this question, the intervention to be effected would change, depending on the result. New therapies could be arranged in order to prevent the possible damages of bradycardia' onset. Studies upon the incidence of Il-1 in respiratory pacemaker are in act to find new therapies to prevent apnea by regulating Il-1 concentration. The exploration of Bi-Channel signals is based on measurements of linear and non-linear relationships between RR and respiratory signals. There are several techniques that allow discovering the relationship between these two channels of signals. Let us briefly introduce these techniques.

*2.2.1. Local linear correlation coefficient ($r^2_{t,f}$).*
It is well known that HRV and respiration are non-stationary. To overcome these difficulties, a new estimator was recently proposed (Ansari-Asl *et al.*, 2005), which aroused our interests. It uses a local linear correlation coefficient, computed the outputs of narrow band-pass filter, as a function of time and frequency, which is maximized according to time delay.

When considering two observations, *x(t)* the HRV signal and *y(t)* the respiration signal, the problem is to characterize the statistical relationship between the nonstationary signals *x(t)* and *y(t)*, simultaneously in the time and frequency domains. The formula of the local linear correlation coefficient is given by:

$$R^2_{x,y}(t,f) = \max_{-\tau_m < \tau < \tau_m} \left( \frac{\left( \sum_{k=-\frac{H}{2}}^{\frac{H}{2}} x_f(k) y_f(k+\tau) \right)^2}{\sum_{k=-\frac{H}{2}}^{\frac{H}{2}} x_f^2(k) \sum_{k=-\frac{H}{2}}^{\frac{H}{2}} y_f^2(k+\tau)} \right) \quad (1)$$

Where $x_f(t)$ and $y_f(t)$ are centered and narrow band filtered signals over a sliding window of duration *H* with an appropriate filter bank (Carrault *et al.*, 2009). $R^2_{x,y}(t,f)$ is calculated with different delays $\tau$ between the two windows. It can be viewed as a set of correlation coefficients computed in different frequency bands when using the Short-Time Fourier Transform (STFT) as continuous filter bank (Ansari-Asl *et al.*, 2005). The parameter $r^2_{t,f}$ is chosen and defined from $R^2_{x,y}(t,f)$ as:

$$r^2_{t,f} = \max_\tau R^2_{x,y}(t,f) \quad (2)$$

*2.2.2. Non-linear Regression coefficient ($h^2$).*
The previous methods suppose a linear relationship between the HRV and the respiration which is not probably acceptable when looking at figure 2. Consequently, we propose one of non-linear indexes, also known as the non-linear regression coefficient, originally used in epilepsy (Pijn *et al.*, 1993).

Non-linear regression coefficient ($h^2$) allows measuring statistical dependence between observations obtained in a bound temporal support. It is defined as:

$$h^2_{X_1 X_2} = \frac{E\left\{ \left( X_2(t) - E\{X_2(t)\} \right)^2 \right\} - E\left\{ \left( X_2(t) - g(X_1(t)) \right)^2 \right\}}{E\left\{ \left( X_2(t) - E\{X_2(t)\} \right)^2 \right\}} \quad (3)$$

$$h^2_{X_1 X_2} = \frac{Var(X_2) - \|X_2 - g(X_1)\|^2}{Var(X_2)} \qquad (4)$$

Where *g* is a non-linear regression function, employed to measure the similarity, more or less linear, between the two observed signals $X_1$ and $X_2$. It is worthwhile to mention the three principal characteristics:

1. For a perfectly linear transformation, the non-linear coefficient $h^2_{X_1 X_2}$ is more and more close to a linear coefficient;
2. $h^2_{X_1 X_2}$ is generally different from $h^2_{X_2 X_1}$, that's why $h^2$ is calculated in both directions;
3. $h^2_{X_1 X_2}$ is a quantity that can assume values between 0 and 1, if $\|X_2 - g(X_1)\|^2 \leq Var(X_2)$

In our application, the non-linear regression coefficient is assessed between the two observations (RR series and Respiration) with several delays, let : $h^2 = \max_\tau h^2_{X_1 X_2}(\tau)$

### 2.3. Methods of Feasibility Study

The long term aim of Feasibility Study is to add a new device to the apnea-bradycardias detection of the used monitors. This new device will be associated with infection detection in NICU. So, in this section, we study the feasibility of its implementation in NICU with the crucial features selected from Mono-Channel analysis and Bi-Channel analysis.

There are two motivations to study feasibility: i) Improve the detection by combining the best features in real time; ii) Propose a pseudo-real time multivariate diagnosis rule based on Optimal Fusion.

For the first motivation, Logistic Regression (LR) is used and it is a useful way of describing the relationship between one or more independent variables (a brief presentation is introduced in the Appendix A). Each of the regression coefficients describes the contribution of its risk factor from two aspects:

a) One aspect is related to Quality: positive vs. negative
   - A positive regression coefficient means that the risk factor increases the probability of sepsis,
   - A negative regression coefficient means that the risk factor decreases the probability of sepsis.

b) The other aspect is Quantity: large value vs. near-zero value
   - A large regression coefficient indicates that the risk factor strongly influences the probability of sepsis,
   - In contrast, a near-zero regression coefficient indicates that the risk factor has little influence on the probability of sepsis.

For the second motivation, a new architecture for the decision making is proposed and based on the combination of several leading features, which is entitled 'Optimal Fusion'.

### 2.3.1. Optimal Fusion.
Different fusion rules can be used to combine local detections (obtained from each algorithm) into a final decision *u*. Some simple rules are based on a "*k* out of *n*" function (*k<n*). Special cases of this function include AND, OR and MAJORITY rules. However, there has been an important effort to obtain the OPTIMAL FUSION rules, often based on the weighted combination of each local detection, which provide a higher weight to the more reliable detectors. In this work we have used the optimality criterion proposed by Chair and Varshney (1986), which has been seen in several studies performed in biomedical field (Hernandez *et al.*, 1999) (Altuve. M. *et al.*, 2011). The principles are briefly recalled here.

Let us consider a binary problem with the two hypotheses as follows:
- $H_0$: sepsis is absent
- $H_1$: sepsis is present.

By denoting a priori probabilities of the two hypotheses: $P(H_0) = P_0$ and $P(H_1) = P_1$, the data fusion rule (Chair and Varshney, 1986) can be expressed by:

$$f(u_1,\cdots,u_n) = \begin{cases} 1, & \text{if } a_0 + \sum_{i=1}^{n} a_i u_i > \lambda \\ -1, & \text{otherwise} \end{cases} \tag{5}$$

As we can observe in (5), the individual detector decisions are formed as a weighted sum and then compared to a threshold $\lambda$. The weights are set below:

$$\begin{aligned} a_0 &= \log \frac{P_1}{P_0} \\ a_i &= \log \frac{1 - P_{M_i}}{P_{F_i}}, \quad \text{if } u_i = +1 \\ a_i &= \log \frac{1 - P_{F_i}}{P_{M_i}}, \quad \text{if } u_i = -1 \end{aligned} \tag{6}$$

Where, $a_0$ is the ratio between a priori probabilities of the two hypotheses;

$a_i$ takes into account either probability of miss or probability of false alarm;

$P_{M_i}$ denotes probability of miss;

$P_{F_i}$ indicates probability of false alarm.

## 3. Results

All recordings were performed in the NICU and physiological time series were recorded in standard conditions. The monitoring (Powerlab system®, AD Instruments) included one-hour recording of two electrocardiogram (ECG), electrooculogram (EOG), electroencephalogram (EEG) leads, one pulse oximetry saturation (SaO2) and nasal respiration trace.

Clinical datasets were obtained from two groups of premature newborns (13 sepsis vs. 13 non-sepsis) hospitalized from the NICU in the Center of Hospital affiliated to University of Rennes 1 (CHU-Rennes) between 2007 and 2010. There were no significant differences in gender, gestational age, chronological age (>72 hours), post-menstrual age (<33 weeks), weight and haematocrit between sepsis and non-sepsis groups. This research was approved by the local ethics committee (03/05-445). Furthermore, the parents of these babies were informed and gave common consents.

Inclusion criteria were more than one bradycardia per hour and/or need for bag-and-mask resuscitation and/or the intention of the attending physician to scrutinize any suspected infection. Whereas, Exclusion criteria implied ongoing inflammatory response with or without confirmed infection, medications known to influence ANS including morphine, catecholamine, sedative drugs, intra-tracheal respiratory support, intra-cerebral lesion or malformation.

Data analysis was executed with signal processing tools designed with MATLAB® R2010b (The Mathworks, Inc.) in Windows® 7. Due to the quasi real-time application, the finite windows adopt 3 sizes listed in table 1, in order to determine which is the best window size for diagnosis.

Table 1. Time duration in finite window.

| Constraint due to the real time:Test on finite window | | | |
|---|---|---|---|
| **Window size** | 1024 | 2048 | 4096 |
| **Time duration (min)** | 4.3 | 8.6 | 17.2 |

Performances were evaluated in view of the receiver operating characteristic (ROC) curves plotted using True Positive rate (TPR) versus the False Positive rate (FPR). By denoting the Area Under Curve (AUC), an infection detector $D_i$ would be considered to be better than an infection detector $D_j$ if $AUC(D_i) > AUC(D_j)$.

### 3.1. Results of Mono-Channel Analysis

The NICU monitoring system acquired ECG signals at 400Hz continuously for one hour, from which the beat-to-beat RR series was constructed and resampled at 4 Hz. The unusual resampling is necessary for premature newborns, which exhibits higher heart rate (120-130 bpm) than adults. These are typical examples of RR series in figure 3. The RR series for Sepsis (left panel) have 18 mins duration, and the other ones (right panel) for Non-Sepsis have 23 mins. Obviously, Sepsis series contains several bradycardias, while Non-Sepsis series have quasi no bradycardias.

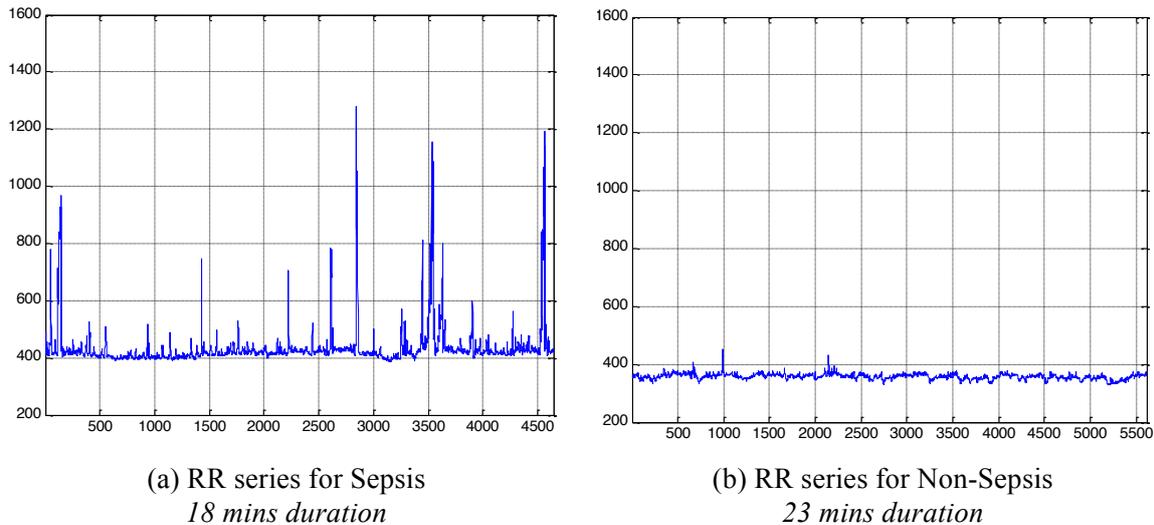

(a) RR series for Sepsis
*18 mins duration*

(b) RR series for Non-Sepsis
*23 mins duration*

**Figure 3.** Typical examples of RR series from babies.

#### 3.1.1. Results from Univariate Analysis.
For this part of results, we conduct the same experiments in 3 sizes of analysis windows (equal to 1024/2048/4096) in order to extract the influential features for sepsis discrimination. Table 2 is the synthesis of results for Univariate Analysis.

Table 2. Univariate Analysis[a].

| | Window Size | | |
|---|---|---|---|
| | 1024 | 2048 | 4096 |
| mean | | | |
| variance | | | |

|  | Window Size | | |
|---|---|---|---|
|  | **1024** | **2048** | **4096** |
| skewness | x | | |
| kurtosis | x | | |
| median | | | |
| SpAs | x | | x |
| SD | | | |
| RMSSD | | | |
| power of HF | | | |
| power of LF | | x | |
| power of VLF | | | |
| $\alpha_{slow}$ (AlphaS) | x | x | x |
| $\alpha_{fast}$ (AlphaF) | x | x | x |
| AppEn | | x | x |
| SamEn | x | x | x |
| PermEn | | | x |
| Regul | | | |

[a]For each case in column, using a ANOVA test, these parameters whose p value less than 0.05 are marked as "x". We use these p values in order to keep a maximum number of significant parameters for the multivariate analysis.

From table 2, certain HRV characteristics such as SD, RMSSD, powers of HF, LF, VLF, were nearly uncorrelated with sepsis. However, results of Univariate Analysis for non-linear methods are very interesting, three indexes from non-linear methods: alphaS, alphaF and SamEn have little p values in all cases. Accordingly, these three can be viewed as candidate parameters to classify sepsis from non-sepsis whatever the size of window. For further details, take window size 1024 as an example, the boxplots in figure 4 display the apparent separation between these two distinct populations.

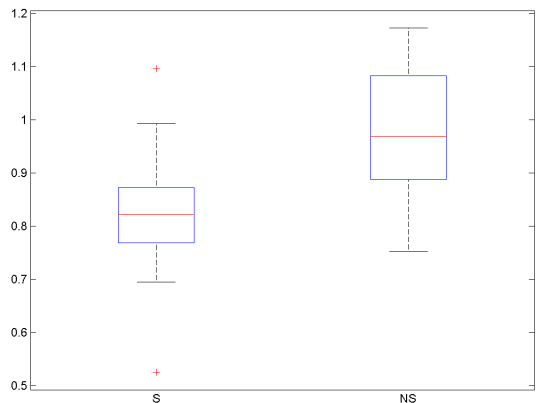
(a) alphaS

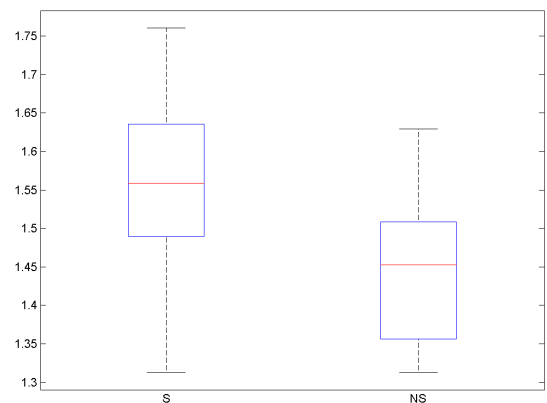
(b) alphaF

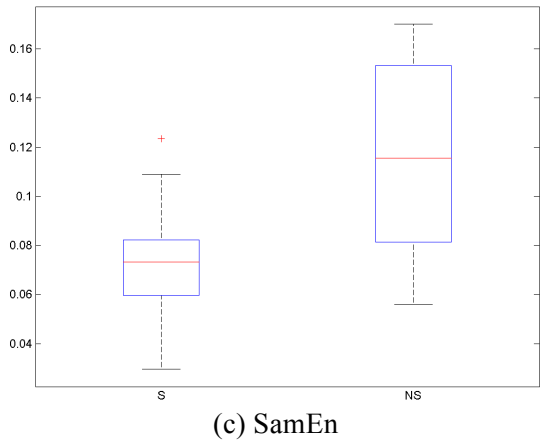

**Figure 4.** Boxplot of candidate parameters for window size 1024.
(a) alphaS, p=0.0012.
(b) alphaF, p=0.0307.
(c) SamEn, p=0.0840.

(c) SamEn

*3.1.2. Results from Multivariate Analysis—Logistic Regression*
In order to keep several relevant parameters, we should focus on the variables that have p-values <0.05. Logistic regression was then used to propose a unique combination of several individual features. Table 3 shows the result of logistic regression, for only one case of size window (1024 samples or in equivalent 4.3 minutes duration).

**Table 3.** Logistic Regression, window size 1024.

| | | | | | |
|---|---|---|---|---|---|
| **Analysis of Maximum Likelihood Estimates** | | | | | |
| **Parameter** | **DF** | **Coefficient** | **Standard Error** | **Wald Chi-Square** | **Pr > ChiSq** |
| **Intercept** | 1 | -12.1413 | 10.5073 | 1.3352 | 0.2479 |
| **SpAs** | 1 | 1.8309 | 0.0576 | 0.7534 | 0.3854 |
| **alphaS** | 1 | -7.4428 | 3.2870 | 5.1273 | **0.0236** |
| **alphaF** | 1 | 9.6729 | 7.2557 | 1.7773 | **0.0182** |
| **SamEn** | 1 | -37.0885 | 25.8890 | 1.6839 | **0.0188** |

Let us briefly explain such a table:
- The first column represents the Intercept and the significant parameters chosen from univariate analysis.
- The second column implies the degree of freedom (DF) of each parameter.
- The third column denotes the estimated coefficients of the parameters respectively, which are $\beta_1, \beta_2, \beta_3, \ldots, \beta_k$ in (7).

$$z = \beta_0 + \beta_1 x_1 + \beta_2 x_2 + \beta_3 x_3 + \cdots + \beta_k x_k \qquad (7)$$

Where:
- z is a measure of the total contribution of all the independent variables used in the model and is known as the logit.
- $\beta_0$ is called the "intercept".
- $\beta_1, \beta_2, \beta_3$, and so forth, are called the "regression coefficients" of $x_1, x_2, x_3$ respectively.
- The fourth column indicates the standard error of the coefficient.
- The fifth column signifies the Wald Chi-Square statistic, computed as the square of the value obtained by dividing the parameter estimate by its standard error.
- The sixth column provides the p-value (Pr > ChiSq) for the Wald Chi-Square statistic with 1 DF, with a value below 0.05 indicating a significant effect of the associated model parameter if a 5 percent significance level is chosen.

From table 3, it is observed that alphaS, alphaF and SamEn have the significant regression coefficients (p-values <0.05). Positive regression coefficients mean that alphaF increases the probability of sepsis, while negative regression coefficients mean that alphaS and SamEn decrease the probability of sepsis. Among these 3 variables, the largest absolute value of regression coefficient means that SamEn strongly influences the probability of sepsis.

We carry out the same analysis for three sizes of window, after, the synthesis of the valuable results is unfolded in table 4, where Significant Regression Coefficients of Logistic Regression are listed.

Table 4. Significant Regression Coefficients of Logistic Regression.

|        | Window Size |          |          |
|--------|-------------|----------|----------|
|        | 1024        | 2048     | 4096     |
| SpAs   | 1.8309      | 0.6450   |          |
| alphaS | -7.4428     | -18.7639 | -11.2460 |
| alphaF | 9.6729      | 16.9458  |          |
| SamEn  | -37.0885    |          | -9.9756  |

From this table, we can see:

Firstly, it is obvious that alphaS is the most frequently chosen as a significant variable with a negative regression coefficient, which decreases the probability of sepsis.

Secondly, alphaF and SamEn are selected for two times.
- Positive regression coefficients indicate that alphaF increases the probability of sepsis.
- On the contrary, negative regression coefficients mean that SamEn decreases the probability of sepsis.

Thirdly, SpAs is also picked for twice, but its absolute values of regression coefficients are below 2, so that we do not need to consider its impact.

### 3.2. Results of Bi-Channel Analysis

During experiments, we calculate $h^2$ between the RR series and raw nasal flux in both directions. Also, we calculate $r^2_{t,f}$ in the same manner of thinking, but in several sub-bands. The same database of patients used for RR series analysis was preserved, adding the recorded respiratory signals. RR series and respiratory signal were first centered horizontally and normalized, due to the fact that the magnitude does not intervene in the computation.

#### 3.2.1. Results from linear method——Time-Frequency plot of $r^2_{t,f}$

In order to attest the idea, figure 5 depicts two typical examples of Time-Frequency plots for $r^2_{t,f}$. The blue signal is the RR series, and the red signal is the nasal respiration trace. Figure 5(a) is the case of non-infected baby, while figure 5(b) corresponds to an infected baby. The Time-Frequency plots of the local correlation coefficient $r^2_{t,f}$ indicate that non-infected group has a strong relationship in the lower band around 0.4 Hz. In contrast, infected group has a higher correlation coefficient in the higher frequency band. These two graphs suggest computing the distribution of the local correlation coefficient greater than 0.8 in different sub-bands.

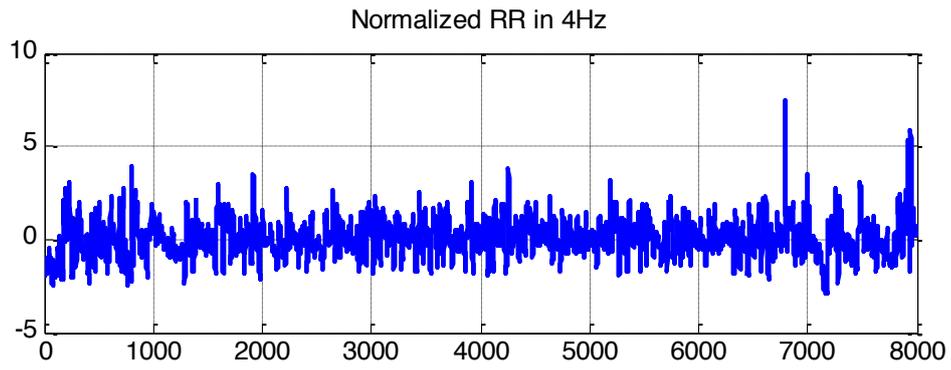
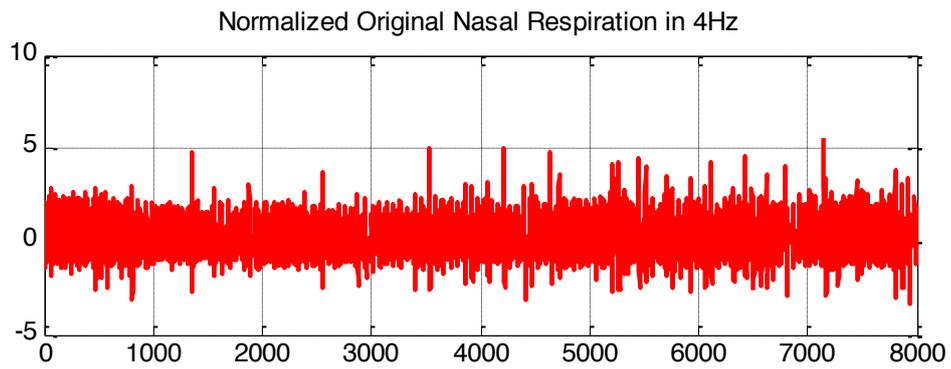
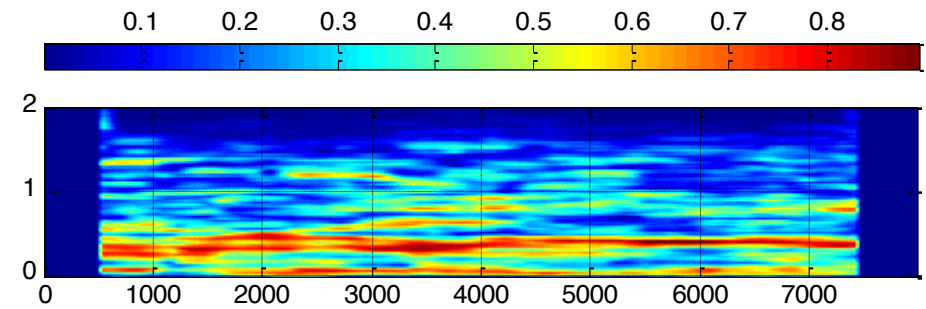

(a) Non-Infected

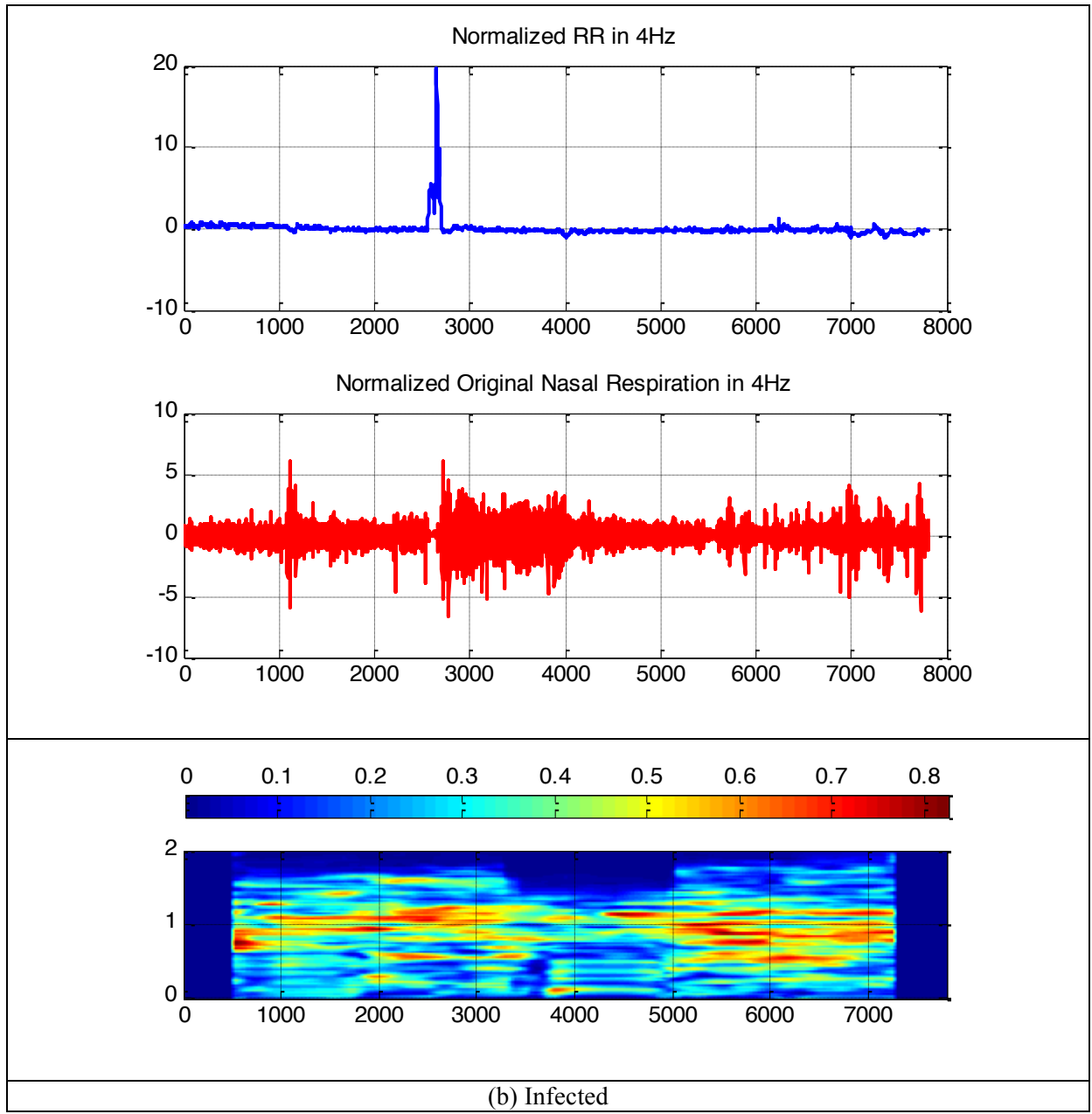

(b) Infected

**Figure 5.** Time-Frequency plot of $r^2_{t,f}$ (a) Non-infected group, we observe a strong relationship in the lower band around 0.4 Hz, (b) Infected group where a higher correlation coefficient in the higher frequency band.

*3.2.2. Results from linear method——Multi Boxplot of $r^2_{t,f}$*
The coming section reports the results of statistical analysis for $r^2_{t,f}$ between centered and normalized RR and nasal respiration frequency band by band. These qualitative findings were statistically verified.
Figure 6 exhibits the distribution of the local linear correlation coefficient $r^2_{t,f}$, greater than 0.8 in different sub-bands.

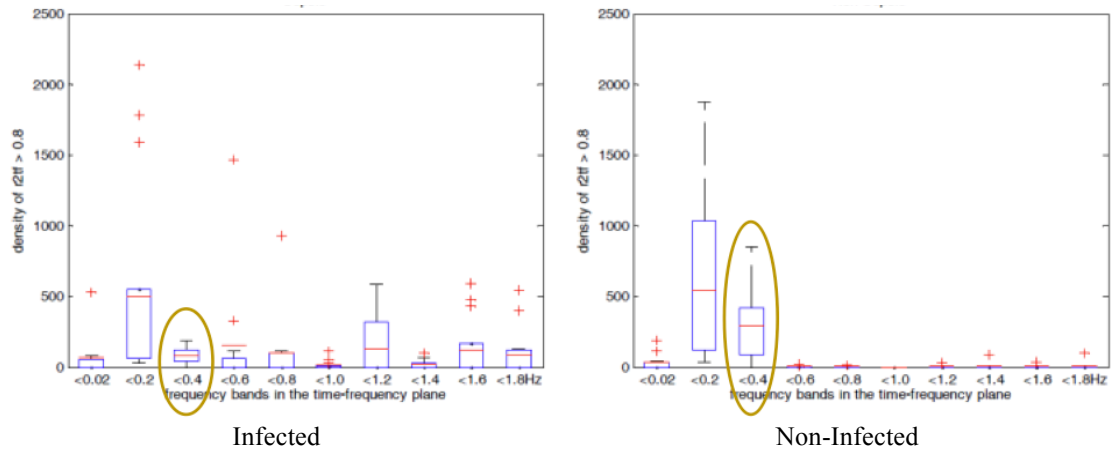

<div style="text-align:center">Infected          Non-Infected</div>

**Figure 6.** Distribution $r^2_{t,f}$ >0.8 in different sub-bands.
In Boxplot, red line is mean value.

Our experiment shows that this difference was significant by the Mann and Whitney (M&W) test between the sepsis group and non-sepsis group within the frequency band 0.2 and 0.4 (circled in brown).

Several experiments were performed and the quantity of $r^2_{t,f}$ between RR and nasal respiration over a threshold set to 0.8 were computed frequency band by band. Table 5 listed here shows the statistical analysis based on M&W test for $r^2_{t,f}$ between RR and nasal respiration in the Window=1024.

**Table 5.** Statistical analysis for $r^2_{t,f}$ Band by Band.

| Band(Hz) | Num Band | Sepsis | Non Sepsis | M&W[a] |
|---|---|---|---|---|
| 0-0.02 | 1 | 7.23 ± 16.68 | 1.38 ± 4.99 | 0.2704 |
| 0.02-0.2 | 2 | 56.07 ±115.25 | 109.53 ±188.12 | 0.6855 |
| **0.2-0.4** | **3** | **8.61 ± 11.33** | **36.23 ± 49.61** | **0.0445** |
| 0.4-0.6 | 4 | 34.53 ± 94.66 | 5.00 ± 8.53 | 0.1881 |
| 0.6-0.8 | 5 | 29.15 ± 53.29 | 72.76 ± 82.25 | 0.1725 |
| **0.8-1.0** | **6** | **1.76 ± 4.32** | **18.53 ± 26.09** | **0.0254** |
| 1.0-1.2 | 7 | 105.61 ±178.86 | 79.23 ±100.97 | 0.5323 |
| 1.2-1.4 | 8 | 10.30 ± 16.19 | 56.84 ± 70.32 | 0.1547 |
| 1.4-1.6 | 9 | 33.61 ± 55.30 | 7.46 ± 11.11 | 0.3446 |
| 1.6-1.8 | 10 | 83.61 ±105.78 | 21.07 ± 37.89 | 0.1995 |

[a]The level of significance was set at p value (p< 0.05)

### 3.2.3. Results from non-linear method—$h^2$

This section presents the results obtained for the non-linear regression coefficient ($h^2$). As an example, applied on the raw data (shown in figure 5), the curves of $h^2(\tau)$ for a sepsis and a non-sepsis with several delays $\tau$ (from 1 to 240) are delineated separately in figure 7 below.

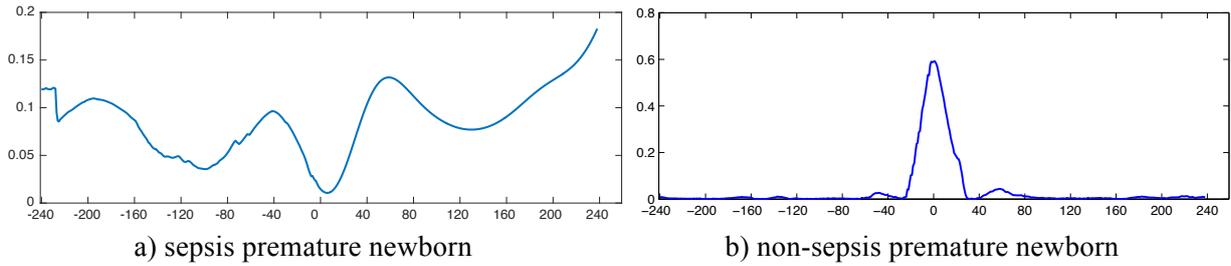

| a) sepsis premature newborn | b) non-sepsis premature newborn |

**Figure 7.** Behavior of the Non-linear Regression coefficient $h^2(\tau)$ for several delays (a) Infected, (b) Non-Infected groups.

From figure 7, it is obvious that the maximum value of $h^2(\tau)$ are very different: non-sepsis baby has a well-defined peak, conversely, sepsis baby has a random shape with no distinct peak.

Non-linear estimates $h^2$ between RR and nasal respiration (h2_rn) and $h^2$ between nasal respiration and RR (h2_nr) were also calculated. The ensuing results are derived from M&W test in the Window=1024.

**Table 6.** Statistical analysis for $h^2$ in two directions.

| h2 | Sepsis | Non Sepsis | M&W[a] |
|---|---|---|---|
| h2_rn | 0.25 ± 0.14 | 0.31 ± 0.17 | **0.03** |
| h2_nr | 0.24 ± 0.16 | 0.33 ± 0.17 | **0.02** |

[a]The level of significance was set at p value (p< 0.05)

*3.2.4. Synthesis of results in Bi-Channel.*
Similar to Mono-Channel analysis, we repeat the experiments of table 5 and table 6 for three sizes of windows 1024/2048/4096, and then summarize all the results in table 7.

**Table 7.** Synthesis of all parameters between RR and nasal respiration.[a]

| | | Window Size | | |
|---|---|---|---|---|
| | | 1024 | 2048 | 4096 |
| $r^2_{t,f}$ | Band(Hz) | | | |
| r2tf1 | 0-0.02 | | | |
| r2tf2 | 0.02-0.2 | | | |
| **r2tf3** | **0.2-0.4** | x | x | x |
| r2tf4 | 0.4-0.6 | | | |
| r2tf5 | 0.6-0.8 | | | |
| r2tf6 | 0.8-1.0 | x | | |
| r2tf7 | 1.0-1.2 | | | |
| r2tf8 | 1.2-1.4 | | | |
| r2tf9 | 1.4-1.6 | | | |
| r2tf10 | 1.6-1.8 | | | |
| $h^2$ | | | | |
| **h2_rn** | | x | x | x |
| **h2_nr** | | x | x | x |

[a]For each case in column, these parameters whose p value less than 0.05 are marked as "×".

The synthesis of results for functional coupling of HRV and nasal respiration shows that the distribution of the local linear correlation coefficient $r^2_{t,f}$, greater than 0.8 in the 3rd frequency band (r2tf3), "h2_rn" and "h2_nr" are the most constantly selected as principal indexes for all 3 window sizes. Thus, these three indicators are regarded as candidate parameters to separate between sepsis and non-sepsis in Bi-Channel.

3.3. Results of Feasibility Study

The target of this section was to study the feasibility of real time detection for sepsis or non-sepsis hypothesis. To reach this aim, firstly, we mix 13 sepsis and 13 non-sepsis infants and then randomly select for 50 times. For each random selection, we connect 26 segments into one long series including 30.3 hours (see figure 8). Therefore, the whole 50 time selections contain 1515 hours.

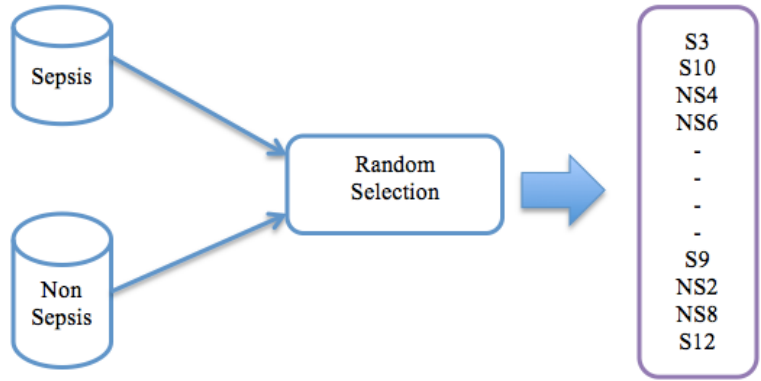

**Figure 8.** Random Selection.

Secondly, on one hand, three features from Mono-Channel analysis: alphaS, alphaF and SamEn were selected. On the other hand, the three estimates r2tf3, h2_rn and h2_nr from Bi-Channel analysis were held as remarkable methods to identify sepsis from non-sepsis. These 6 candidate parameters are recalculated over these 50 long series in the sliding window with the same length step using three sizes 1024/2048/4096. Here, we count each window of each long series as an instance. In other words, each window is viewed as a unique patient. Finally, we performed the Five tests summarized in table 8 for these long series.

Table 8. Five tests for Feasibility Study.

| Test | Explanations |
|---|---|
| Test 1 | Plot 50 ROC curves for 3 features from Mono-Channel: alphaS, alphaF and SamEn. |
| Test 2 | Optimal Fusion for these 3 features in Mono-Channel. |
| Test 3 | Plot 50 ROC curves for 3 features from Bi-Channel: r2tf3, h2_rn and h2_nr. |
| Test 4 | Optimal Fusion for these 3 features in Bi-Channel. |
| Test 5 | Optimal Fusion for the aforementioned 6 candidate parameters. |

3.3.1. Test 1.
In section 3.1., alphaS, alphaF and SamEn are identified as eminent methods to segregate sepsis from non-sepsis, and the ROC curves computed 50 times selections over three different sizes of window. In this case, the procedure for decision making is to compare the statistic $S(x)$ to thresholds,

let
$$S(x) > \lambda \qquad (8)$$
where *S(x)* is either alphaS, alphaF or SamEn.

Take alphaS as an example. For each long series, we compare alphaS values with 1000 thresholds and then plot ROC curves for 50 random selections in figure 9. Furthermore, the red curve in figure 10 illustrates the best ROC curve, which is nearest-upper-left and on which the red point is the closest to upper left corner (0,1). Here, these values of FPR and TPR are used as weights of data fusion rule.

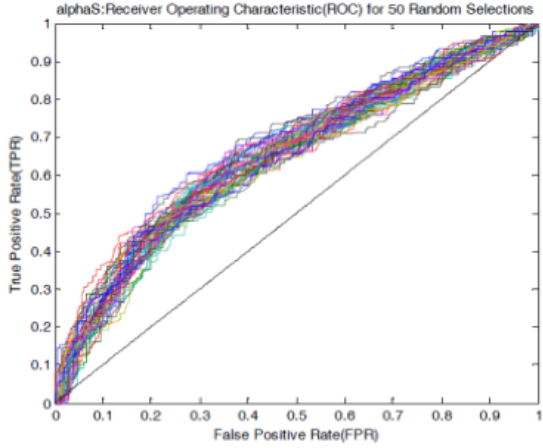
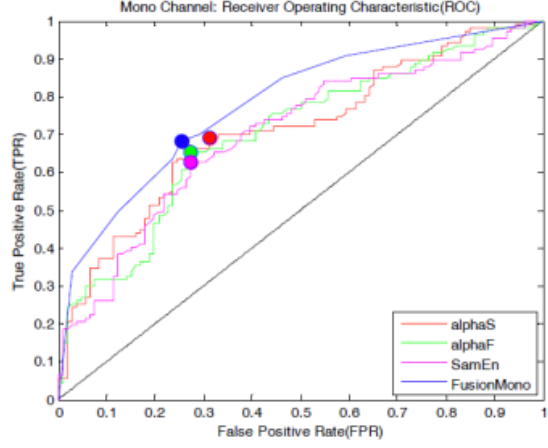

**Figure 9.** ROC curves for 50 random selections in Test 1 (alphaS as an example).

**Figure 10.** The nearest-upper-left ROC curves, where the points the closest to upper left corner (0, 1) are marked as colourful round points.

*3.3.2. Test 2.*
We combined alphaS, alphaF and SamEn using Optimal Fusion in Mono-Channel (FusionMono), and then the best ROC curve of FusionMono is illustrated in figure 10 (blue curve). The best ROC curve of each nominee feature in Mono-Channel from Test 1 and the best ROC curve of FusionMono from Test 2 are drawn together in figure 10.

Compare the ROC curves of alphaS, alphaF, SamEn, and FusionMono, we observe that i) Optimal Fusion Mono has larger AUC than any single of 3 features from Mono-Channel; ii) It improves performance in Mono-Channel.

*3.3.3. Test 3.*
In section 3.2., r2tf3, h2_rn and h2_nr are recognized as outstanding methods to recognize sepsis from non-sepsis. In Test 3, there are two cases:
- In the case of $r^2_{t,f}$, it counts the number of time that the value of $r^2_{t,f} > 0.8$ within frequency band (*0.2<f<0.4 Hz*), and then the statistics *S(x)* compare this number to a threshold. The final statistic is
$$S(x) = N_{0.2}^{0.4} > \lambda \qquad (9)$$
- In the case of $h^2$, the statistic *S(x)* compares the values of $h^2$ to the thresholds, let:
$$S(x) = h^2 > \lambda \qquad (10)$$

The experimental condition of the previous section was re-conducted here, but using two channels (RR series and respiration). Figure 11 shows 50 ROC curves of h2_rn as an example. Moreover, figure 12 shows its best ROC curve (green curve), which is nearest-upper-left and on which the green point is the closest to upper left corner (0,1). As previously mentioned, these values of FPR and TPR are used as weights of data fusion rule for the measurements in Bi-Channel.

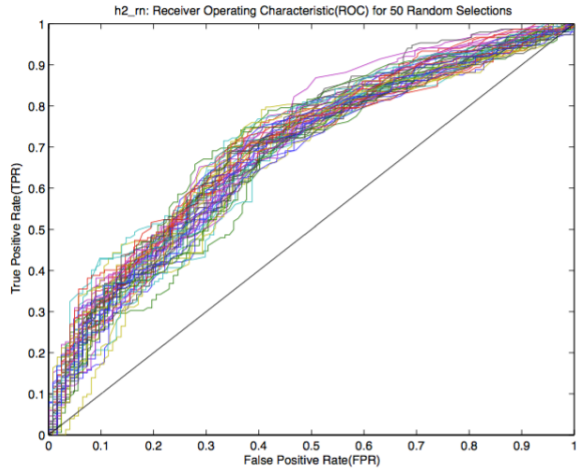
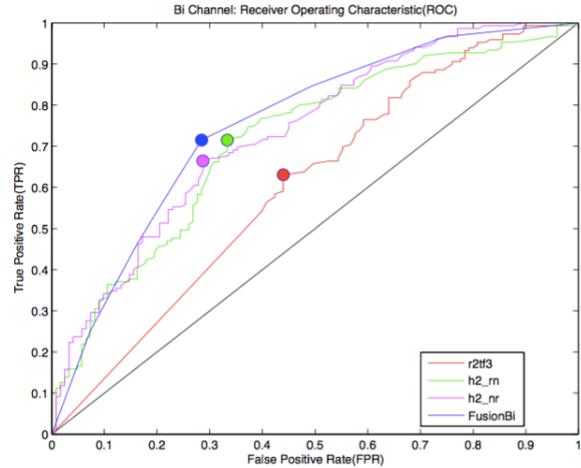

**Figure 11.** ROC curves for 50 random selections in Test 3 (h2_rn as an example).

**Figure 12.** The nearest-upper-left ROC curves, where the points the closest to upper left corner (0, 1) are marked as colourful round points.

*3.3.4. Test 4.*
We blended r2tf3, h2_rn and h2_nr using Optimal Fusion in Bi-Channel (FusionBi), after, the best ROC curve of FusionBi is demonstrated in figure 12 (blue curve). The nearest-upper-left ROC curve of each candidate feature in Bi-Channel from Test 3 and the best ROC curve of FusionBi from Test 4 are collectively illustrated in figure 12.

Compare the ROC curves of these three indexes and FusionBi, we observe that: i) Optimal Fusion Bi has larger AUC than any single of 3 features from Bi-Channel and ii) It also improves performance in Bi-Channel.

*3.3.5. Test 5.*
Test 5 mixed all noteworthy features in Mono-Channel and Bi-Channel using Optimal Fusion. Furthermore, the important characteristics of ROC curves analysis are reported in table 9 in order to dig out the best method for sepsis diagnosis. All the results are based on 5-fold cross validation. That is to say, we divide all instances of 50 long series into 5 subsets. Each algorithm is evaluated on each subset with all other 4 subsets being the training dataset. The final result is the mean of all evaluated subsets. The following is the explanations of table 9:
- The first column indicates the type of test.
- The second column lists features of each test.
- The third column delivers FPR values, which represent probability of false alarm ($P_{FA}$).
- The fourth column provides TPR values, which denote probability of detection ($P_D$).
- The fifth column presents the Area Under Curve (AUC), which is the most important criteria to compare among all features. The larger the AUC is, the more convincing the decision-making method is.
- The best performance is in bold in table 9.

Table 9. Characteristics of ROC curves, window size 1024.

| Test | Features | FPR($P_{FA}$) | TPR($P_D$) | AUC |
|---|---|---|---|---|
| Test 1 | alphaS | 0.2783 | 0.6075 | 0.6630 |
|  | alphaF | 0.2919 | 0.6175 | 0.6684 |
|  | SamEn | 0.3714 | 0.6759 | 0.6648 |
| Test 2 | Optimal Fusion Mono-channel | 0.3019 | 0.6636 | 0.4191 |

| Test | Features | FPR($P_{FA}$) | TPR($P_D$) | AUC |
|---|---|---|---|---|
| Test 3 | r2tf3 | 0.4400 | 0.6309 | 0.4885 |
|  | h2_rn | 0.3333 | 0.7152 | 0.7185 |
|  | h2_nr | 0.2869 | 0.6645 | 0.7359 |
| Test 4 | Optimal Fusion Bi-channel | 0.2846 | 0.7152 | 0.5557 |
| Test 5 | Optimal Fusion All | 0.2149 | 0.7516 | **0.7731** |

From table 9, it is evident that Optimal Fusion all (FusionAll):
- has larger AUC than any single of 6 features from Mono-Channel and Bi-Channel.
- has larger AUC than FusionMono and FusionBi.
- definitely improves performance than any other fusion strategy.

In others words, Test 5 has the least $P_{FA}$ and the largest AUC, consequently, Test 5 is considered as a new methodology.

## 4. Discussion

### 4.1. Discussion of Mono-Channel Analysis

The aim of the first part of our work was achieved by RR signal analysis. With HRV characteristics such as quantitative estimates in Time Domain as well as Frequency Domain, we were unable to find a correlation between these parameters and sepsis. Hence, chaos indexes (alphaS, alphaF) and four metrics from Information Theory were considered (AppEn, SamEn, PermEn and Regul).

Referring to Chaos Theory, on one hand, the index alphaF obtained by DFA and characterizing short-range (4–40 beats) correlation in the detrended RR time series hardly changed with age and was almost constant at about 1.5 in the mean. The alphaF value greater than unity in the preterm infants indicated that the RR fluctuation at short range was near to Brownian motion (uncorrelated). On the other hand, the index alphaS characterizing the long-range (40–1000 beats) correlation increased, and showed the high correlation coefficient value and most statistically significance between sepsis and non-sepsis groups, suggesting that alphaS could be a good and robust index characterizing the ANS development.

With regard to Information Theory, results confirmed the relationship between the occurrence of disease and a reduction of information carried by cardiovascular signals. AppEn, SamEn and PermEn showed that a decrease of entropy is associated with sepsis condition. Conversely, the Regul index measured a higher value for the same group of patients.

Furthermore, all methods are screened through statistical analysis. Finally, three indexes from non-linear methods ---- alphaS, alphaF and SamEn are selected as candidate parameters from Mono-Channel Analysis to differentiate between the two groups—Sepsis vs. Non-Sepsis, because these three are constantly significant whatever size of window. Thus, the distinctive variation in heart rate behavior connected with sepsis could be useful in the field of neonatology.

### 4.2. Discussion of Bi-Channel Analysis

The second part of work was based on the functional coupling of HRV and respiration. We use the identical patients as Mono-Channel analysis, coupled with their respective nasal flux traces. Both linear and non-linear relationships have been measured. Linear method was time-frequency index ($r^2_{tf}$), while a non-linear regression coefficient ($h^2$) was used to analyze non-linear relationships.

Concerning linear estimates, we confirm statistically ($p<0.05$) that the higher correlation is retrieved in the low frequency band for the sepsis group between RR and nasal respiration. Results show that the relationships are circumscribed within a specific region of the 3$^{rd}$ time-frequency plane ($0.2<f<0.4$ Hz) or r2tf3 are different between sepsis and non-sepsis, no matter what size of window. In addition, several $r^2_{tf}$ thresholds ($0.6, 0.7, 0.8, 0.9$) were tested and the threshold $0.8$ appears the most discriminative between the two groups.

Regarding non-linear measures, having such a clinically interesting background, accordingly they are used as mathematical and statistical tools to discriminate the two categories of sepsis and non-sepsis signals. Regardless of window size, the result of $h^2$ as non-linear regression coefficient between RR and nasal respiration in two directions (h2_rn, h2_nr) were always significant to sort out the two groups during the whole process of statistical analysis. Furthermore, the graphs of non-linear relationship show that the $h^2$ function presents a well-defined peak for non-sepsis case, on the contrary, an arbitrary shape in sepsis group is often observed. The shift of the $h^2$ coefficient may also indicate an alteration of the stretch receptor function due to the infection.

*4.3. Discussion of Feasibility Study*

As for the third part of our research, we implement feasibility study on the candidate parameters selected from Mono-Channel Analysis and Bi-Channel Analysis respectively, and their mixed condition. Among all tests, the proposed Test 5 based on optimal fusion of all 6 features (alphaS, alphaF, SamEn, r2tf3, h2_rn and h2_nr) shows good performance with the least $P_{FA}$ and the largest AUC. In addition, the contrast among the three window sizes 1024/2048/4096 is reported in table 10.

Table 10. Contrast among the three window sizes in Test 5.

| Window Size | minutes | FPR($P_{FA}$) | TPR($P_D$) | AUC |
|---|---|---|---|---|
| **1024** | 4.3 | 0.2149 | 0.7516 | **0.7731** |
| **2048** | 8.6 | 0.1803 | 0.7895 | **0.7813** |
| **4096** | 17.2 | 0.1000 | 0.8421 | **0.8246** |

This table indicates that Window 4096 or taking a decision every 17.2 mins has the least $P_{FA}$ and the largest value of AUC.

## 5. Conclusions

The objective of our research work is to determine if heart rate variability (HRV), respiration and their relationships help to diagnose infection in NICU monitoring system via non-invasive ways. To realize our goals, we demonstrate three parts of results:

First of all, we studied merely the RR interval series. Many of our features are inspired by classic signal analysis methods including RR series distribution patterns (mean, variance, skewness, kurtosis, median, SpAs), magnitude of variability in the time domain (SD, RMSSD), and linear estimates in the frequency domain (power of VLF, LF, HF).

Besides, we also make use of features kindled by chaos theory and information theory, which are helpful in analyzing the degree of self-affinity and randomness of the time-series. Specifically, we adopt the detrended fluctuation analysis (DFA) indexes, such as alphaS and alphaF, based on chaos theory to assess the statistical self-similarity of a signal. We also compute the randomness estimates including AppEn, SamEn, PermEn and Regul. They are measurements designed to quantify the degree of regularity versus unpredictability, reflecting the unpredictability of fluctuation in a signal. A low value of the entropy indicates that the signal is deterministic, whereas a high value means that the signal is unpredictable. These entropies are good indicators for cardiovascular signals where the occurrence of sepsis is highly correlated with the decrease of entropy.

For each method, we attempt three sizes of window 1024/2048/4096, and then compare these methods in order to find the prominent means to distinguish sepsis premature infants from non-sepsis ones. The results show that **alphaS**, **alphaF** and **SamEn** are essential parameters to recognize sepsis from the diagnosis of late onset infection in premature infants.

However, in sick premature infants, the mechanism is probably not just a change in RR series. The clinical findings here clearly demonstrate that HRV, respiration and their relationship could be efficient diagnosis tools and may help identifying culture-positive sepsis in a population of infants with unusual and recurrent apnea-bradycardia. The patients used for RR analysis were retained, joining respiratory signals for the same cohort.

Second, the question about the functional coupling of heart rate variability and nasal respiration was also addressed. We studied the linear time-frequency index ($r^2_{t,f}$), as well as a non-linear regression coefficient ($h^2$). In particular, we considered the two directions of coupling during estimate the index $h^2$ of non-linear regression. Finally, from the entire analysis process, it is obvious that the three indexes as follows:

- the quantity of $r^2_{t,f}$ between RR and nasal respiration over a threshold set to 0.8 in the 3$^{rd}$ sub-band *0.2<f<0.4 Hz* (**r2tf3**)
- $h^2$ between RR and nasal respiration (**h2_rn**)
- $h^2$ between nasal respiration and RR (**h2_nr**)

were complementary methods to diagnose sepsis in such delicate patients by a non-invasive way.

Third, feasibility study is carried out on the candidate parameters selected from Mono-Channel Analysis and Bi-Channel Analysis respectively. Firstly, we generate long series mixing sepsis and non-sepsis cases as real time series. Next, we test the sepsis or non-sepsis hypothesis on every segment of 3 kinds of window size by 5 types of Test. Here, an optimal fusion strategy is proposed and based on the mixed condition of significant features. Finally, we evaluate the characteristics of ROC curves such as $P_{FA}$, $P_D$ and AUC. After comparing all these tests, we discovered that the proposed Test 5 based on optimal fusion of 6 features (alphaS, alphaF, SamEn, r2tf3, h2_rn and h2_nr) shows good performance with the least $P_{FA}$ and the largest AUC, which can be used to provide high-precision warning alerts of apnea-bradycardia in NICU monitoring system. In addition, the comparison among the three window sizes indicates that Window 4096 has the least $P_{FA}$ and the largest value of AUC.

As a conclusion, we believe that the selected measures from Mono-Channel and Bi-Channel signal analysis have a good repeatability and accuracy to test for the diagnosis of sepsis via non-invasive NICU monitoring system, which can reliably confirm or refute the diagnosis of infection at an early stage.

**Appendix A. Logistic Regression**

In statistics, logistic regression (sometimes called the logistic model or logit model) is used for predicting the probability of occurrence of an event by fitting data to a logistic curve. It is a Generalized Linear Model (GLM) used for binomial regression. Like many forms of regression analysis, it makes use of several predictor variables that may be either numerical or categorical.

An explanation of logistic regression begins with a logistic function, which always takes on values between zero and one:

$$f(z) = \frac{e^z}{e^z + 1} = \frac{1}{1 + e^{-z}} \tag{A.1}$$

The input is z and the output is *f*(z). The logistic function is useful because it can take as an input any value from negative infinity to positive infinity, whereas the output is confined to values between 0 and 1. The variable z represents the exposure to some sets of independent variables, while *f*(z) denotes the probability of a particular outcome, given a set of explanatory variables. The variable z is a measure of the total contribution of all the independent variables used in the model and is known as the logit.

The variable z is usually defined as

$$z = \beta_0 + \beta_1 x_1 + \beta_2 x_2 + \beta_3 x_3 + \cdots + \beta_k x_k \tag{A.2}$$

Where
  $\beta_0$ is called the "intercept", which is the value of *z* when all independent variables are zeros.
    (e.g. the value of *z* in someone with no risk factors).
  $\beta_i$ (*i*=1 to *k*) is called the "regression coefficient" of $x_i$ respectively.

Each of the regression coefficients describes the contribution of risk factor. As for quality, a positive regression coefficient means that the explanatory variable increases the probability of outcome, while a negative regression coefficient means that the variable decreases the probability of outcome. In terms of quantity, a large regression coefficient means that the risk factor strongly influences the probability of outcome, whereas a near-zero regression coefficient means that the risk factor has little influence on the probability of outcome. In our paper, we regard sepsis as outcome.

Logistic regression is a useful way to describe the relationship between independent variables (e.g., age, sex, etc.) and a binary response of classification, which has only two categories, for example, "has sepsis" or "doesn't have sepsis".